\input harvmac

\input epsf
\ifx\epsfbox\UnDeFiNeD\message{(NO epsf.tex, FIGURES WILL BE
IGNORED)}
\def\figin#1{\vskip2in}
\else\message{(FIGURES WILL BE INCLUDED)}\def\figin#1{#1}\fi
\def\ifig#1#2#3{\xdef#1{fig.~\the\figno}
\goodbreak\topinsert\figin{\centerline{#3}}%
\smallskip\centerline{\vbox{\baselineskip12pt
\advance\hsize by -1truein\noindent{\bf Fig.~\the\figno:} #2}}
\bigskip\endinsert\global\advance\figno by1}

\def\cst {{\rm const.}}

\def \ov {\over}

\def \lr { \lref}

\def\dd {\partial }

\def\tt{{\cal T}(V)}

\def\n{\noindent}
\gdef \jnl#1, #2, #3, 1#4#5#6{ { #1~}{ #2} (1#4#5#6) #3}
\def\epsi{ {\cal E} }

\def\np {  Nucl. Phys. }
\def \pl { Phys. Lett. }

\def \prl { Phys. Rev. Lett. }
\def \pr  { Phys. Rev. }
\def \cqg { Class. Quant. Grav. }

\lr \beken {J.D. Bekenstein, \pr D7 (1973) 2333; D9 (1974) 3292.  }
\lr \hawk {S. Hawking, Commun. Math. Phys. 43 (1975) 199.  }
\lr \thoof { G. 't Hooft,  \np B256 (1985) 727;
  \np B335 (1990) 138.  }
\lr \hooft { G. 't Hooft, Physica Scripta T36 (1991) 247; 
 {\it Dimensional reduction in quantum gravity},
Utrecht preprint THU-93/26, gr-qc/9310006. }
\lr\hft{C. Stephens, G. 't Hooft and B.F. Whiting,
\cqg 11 (1994) 621. }
\lr \sussk {L. Susskind,  {\it The world as a hologram},
  preprint SU-ITP-94-33, hep-th/9409089. }
\lr\ssk{ L. Susskind, \prl 71 (1993) 2367. } 
  \lr \rst {J.G. Russo, L. Susskind and L. Thorlacius,  
\pr D46 (1992) 3444; \pr D47 (1993) 533.  }
\lr \stu {L. Susskind, L. Thorlacius and J. Uglum,  
\pr D48 (1993) 3743.  }
\lr \membr {K. Thorne, R. Price and D. MacDonald, {\it Black holes:
the membrane paradigm} (Yale Univ. Press, New Haven, CT, 1986). }
\lr \rstf  {J.G. Russo, L. Susskind and L. Thorlacius, \pl B292 (1992) 13.}  
\lr \hori {J.G. Russo, \pl B359 (1995) 69.} 
\lr\wald {R. M. Wald, {\it General Relativity} (University of
Chicago Press, Chicago, 1984).}
 \lr \ssvv {K. Schoutens, H. Verlinde and E. Verlinde,
 preprint CERN-TH.7142/94, hep-th/9401081.}
 \lr \svv {Y. Kiem, H. Verlinde and E. Verlinde, \pr D52 (1995) 7053.} 
 \lr\giddings{S. Giddings, \pr D46 (1992) 1347.}
\lr\jacob {T. Jacobson, \pr D44 (1991) 1731.}
\lr\jacobs{T. Jacobson, preprint
Utrecht THU-96/01, hep-th/9601064. }

\lr\amru{D. Amati and J.G. Russo, {\it Black holes by analytic
continuation}, hep-th/9602125. }

\baselineskip8pt
\Title{\vbox
{\baselineskip 6pt{\hbox{CERN-TH/96-33}} {\hbox{hep-th/9602124}} {\hbox{
   }}} }
{\vbox{\centerline {Model of black hole evolution }
  }}
\vskip -20 true pt

\centerline  { J.G. Russo
 }

 \smallskip \bigskip
 
\centerline{\it  Theory Division, CERN}
\smallskip

\centerline{\it  CH-1211  Geneva 23, Switzerland}

\bigskip\bigskip\bigskip
\centerline {\bf Abstract}
\bigskip

From the   postulate that  a black hole can be
replaced by a boundary on the apparent horizon with suitable boundary
conditions, an unconventional scenario for the evolution emerges.
Only an insignificant fraction of energy of order $(mG)^{-1}$ is radiated
out. The outgoing wave carries a very small part of the quantum 
mechanical information of the collapsed body, the bulk of the information remaining in the final stable  black hole
geometry.

\medskip
\baselineskip8pt
\noindent

\Date {February 1996}
\noblackbox
\baselineskip 14pt plus 2pt minus 2pt

\vfill\eject
 
It has been argued that, due to back-reaction effects, the Hawking model of
black hole evaporation \hawk\ may break down long before the evaporation is complete \refs {\thoof, \jacob }.
Because of the exponential redshift, the outgoing modes   arise
   from a reservoir of transplanckian 
energies, with frequencies even higher than the total black hole
mass. 
If a Planck-scale cutoff is imposed before the horizon,  
 it seems that there would     be only a scarce amount of     outgoing modes,   and black holes
would   lose an insignificant mass by evaporation   \jacob . 
 Lacking the fundamental short-distance theory,
by the time the outgoing modes arise with Planck frequencies, some extra assumption is needed. Extrapolating the Hawking radiation into this region
leads to paradoxes, e.g.   loss of quantum coherence.  
However, a concrete alternative scenario to the Hawking model has been   elusive so far.

Recently, there have been some indications on how  the problem should be formulated \refs{\stu , \hft }.
The   idea is that the   description of    physics which is appropriate to external observers may require imposing a phenomenological boundary 
on a surface (the `stretched' horizon \membr ) located  about one Planck unit away from the event horizon,  where gravitational self-interactions become very strong   \stu  .
It can be seen, in particular, that  the apparent horizon is always
inside the stretched horizon and it coincides
with it once the supply of  collapsing energy-density flux 
is over \hori.
 
In this letter a novel theory of black hole evaporation
 will be constructed and examined.  It will be assumed that the adequate   framework for
outside observers is based on a quantum   theory with a boundary
on the apparent horizon.
The outgoing flux of energy  in this model
will    coincide with the one predicted by the Hawking model
 only in the region which is not causally connected with
the apparent horizon. By that retarded time the Hawking radiation flux is still negligible.
In the region in causal contact with the boundary the total flux will be 
very small and it will exponentially go to zero.
As a result, the final state will contain a stable geometry with approximately the same mass as the ADM mass of the original configuration. Only an energy of
order $(mG)^{-1} $  will be evaporated.\foot{
There is another well-known disturbing feature of the Hawking
model, namely that at the endpoint of the evaporation the curvature singularity
remains exposed to outside observers.
This problem is thus absent in the model described here.}

 A scenario where the Hawking radiation stops leaving a macroscopic
black hole was contemplated by Giddings as a possible solution of the information problem.
In ref. \giddings\ it was suggested that the radiation should stop when
a certain bound on the information content is saturated.
 Although this is not what the present model predicts, the way
the  information paradox is resolved is similar,
the Hawking process terminates and  the information remains stored in the final black hole  geometry.
  


Let us   restrict our attention to spherically symmetric configurations,
  \eqn\bmet{
ds^2=g_{ij}(x^0,x^1) dx^idx^j+  r^2(x^0,x^1) d\Omega ^2
\ , \ \ \ \ i,j=0,1\ .
 }
 In this spherically symmetric space-time,  the location of the apparent horizon
  is determined by $g^{ij}\dd_i r \dd_j r=0\ $ (see e.g. ref. \hori ).
 In the conformal gauge, $g_{ij}(x^0,x^1) dx^idx^j=e^{2\rho (U,V)} dUdV$,
 this equation takes the form $ \dd_U r\dd_V r=0 $.
For the part of $U,V$ space which is physically relevant in the process of
  gravitational collapse,
the apparent horizon will be simply given by the 
equation $\dd_V r=0$.
Below we will first determine the apparent horizon curve in terms of the
infalling matter, and then calculate the outgoing energy-density fluxes
by implementing suitable boundary conditions 
on the apparent horizon. 


 Let $r=r(U,V), \ \rho=\rho (U,V)$ be the classical solution of the Einstein equations
for a given infalling  spherically symmetric configuration
$T_{\mu\nu} $. For simplicity  only   massless matter will be considered.
In the conformal gauge, the classical Einstein equations for the $g_{UU}$ and
$g_{VV}$ components  are given by
\eqn\tein{
\dd_U ^2 r -2 \dd_U \rho  \dd_U r=- 4\pi G r  T_{UU} \ ,\ \ 
 \ \  
\dd_V ^2 r -2 \dd_V \rho  \dd_V r= - 4\pi G r  T_{VV} \ ,
}
where $ T_{VV}$ and $T_{UU}$ represent   incoming and outgoing energy-density fluxes. Let the apparent horizon curve be classically given by $U=-P(V)$.
It is easy to obtain $r(U,V)$ in the neighborhood of the apparent horizon. 
Expanding around $U=-P(V)$, we have $\dd_V r^2\equiv f(U,V)= -F(V) (U+P(V))+
O\big( (U+P(V))^2\big) $. By   a conformal transformation one can always set $F(V)dV\to {\rm const.} dV$ ,  so that   the equation simply becomes 
 $\dd_V r ^2=  -{\rm const. } (U+P(V))+...$  It is convenient to choose the multiplicative constant equal to $2e^{-1}$  (cf. eqs. (6), (8)). 
By integration
  we obtain
\eqn\rrrq{
r^2(U,V)= (2M(V)G)^2 - 2e^{-1} V(U+P(V))+ O\big((U+P(V))^2 \big)\ ,
}
where a possible additive function $f(U)=c (U+P(V))+...$ is removed by a shift of $V$,
and  a function $M(V)$ was introduced, defined by  
\eqn\mmm{
2eG^2 {dM^2(V)\ov dV} = V {dP(V)\ov dV}\ .
}
Using eqs. \tein ,   \rrrq\ and \mmm ,  the functions $M(V), P(V)$ can be related to the
incoming energy momentum tensor.  In particular, evaluating the $VV$-constraint \tein\ near the apparent horizon, the second term can be dropped, and one
 finds  
\eqn\ppmm{
 {dP(V)\ov dV}\cong    {T_{VV}\ov \tt }\ ,\ \ \ \ \tt \equiv
\big(16\pi e G^3 M^2(V) \big)^{-1}  \ .
}
To fix the notation, let us consider the static Schwarzschild geometry. The  
 standard
connection with Kruskal coordinates $U,V$ is given by  
\eqn\trece{
2mG (r-2mG)e^{r\ov 2mG}=-V(U+p)\ , \ \ \ \ p=2mG\ ,
}
\eqn\uuvv{
U+p=-2mG e^{-{u\ov 4mG}}\ ,\ \ \ V=2mG e^{{v\ov 4mG}}\ , \ \ \ 
v,u=t \pm r^*\ ,\ \ r^*=r  + 2mG \log (r-2mG)\ .
}
In this case the apparent horizon   coincides with the event horizon. The solution of $\dd_V r=0$ is $U=-p$. Expanding $r(U,V)$ in eq. \trece\ 
near $U=-p$ one obtains
\eqn\itera{
r ^2\cong (2m G)^2 - 2e^{-1} V(U+p)  +O\big(  (U+p)^2 \big)\ .
} 
For a dynamically formed black hole, assuming that $T_{VV}$ vanishes
for $V>V_1$, $m=M(V_1)$ will represent
the total ADM mass of the collapsing body, and $p=P(V_1)$ will be associated with the total infalling Kruskal momentum.

The equation of the apparent horizon in the absence of incoming fluxes
was determined in  \hori . It is easy to generalize this
calculation to incorporate infalling matter.
 The equation ${\dd r(U,V)\ov \dd V}=0$  can be written in terms of the
total derivative on the apparent horizon curve, 
$r=r_{\rm AH}\big( U, V(U)\big)$,
\eqn\aapp{
0={d r_{\rm AH}\ov d U}-{\dd r_{\rm AH}\ov \dd U}\cong 
{d r_{\rm AH}\ov d U}+{1\ov 2eMG} V 
\ .
}
For a large Schwarzschild black hole, $r_{\rm AH}\cong 2MG$, so that
$ -{V\ov 2eMG}={dr_{\rm AH}\ov dU} \cong 2G {dM\ov dU}\ .$
 In the
vicinity of the horizon, a black hole loses mass at a rate as dictated
by the Stefan-Boltzmann law, and it gains mass in accordance to
the incoming energy-density flux,   
\eqn\docec{
{d M\ov d  v}=\bigg(  -N{\pi^2\ov 30} T^4_H  + T_{vv} \bigg) (4\pi r_s^2) \  ,
\ \ \ \ \ r_s=2 M(V)G\ ,
}
where $N$ represents the number of scalar field  degrees
of freedom.   
Using ${dM\ov dU} = {dM\ov dv} {dv\ov dV}{dV(U)\ov dU}$,
and eqs. \uuvv , \aapp ,   \docec , one obtains
\eqn\doce{
{d V\ov d  U}\bigg[ -{N e G m^2 \ov 480 \pi M^2 V^2 } + {T_{VV} \ov \tt }\bigg] =-1
\ .\ \ 
}
From eq. \doce\ (see also eq. \docec ) we see that there is a critical value   of the incident
energy-density flux      for which $dV/dU $ changes sign: for lower
$T_{VV}$ the apparent horizon is time-like; for larger $T_{VV}$, 
it is space-like. 
Note that a space-like apparent horizon necessarily involves a black hole geometry, since it implies that the curve $r(U,V)=0$ is space-like.
In Minkowski coordinates:
\eqn\tcrit{
T_{vv}^{\rm cr }\bigg|_{\rm AH} = N{\pi^2\ov 30} T^4_H={N\ov 122880 \pi ^2 G^4 M^4}  \ . } 
Equation \doce\ can be easily integrated when $V$ is   close to $V_1$, where $ M(V)\cong m$.
In this region the apparent horizon curve takes the simple form
\eqn\gahh{
V(U+ P(V))\cong -kG\ ,\ \ \ \ \ \ \ k={Ne\ov 480\pi }\ .
}


\ifig\fone{ 
Apparent horizon   for an   incident flux   less than critical.
}{\epsfxsize=5.6cm \epsfysize=5.1cm \epsfbox{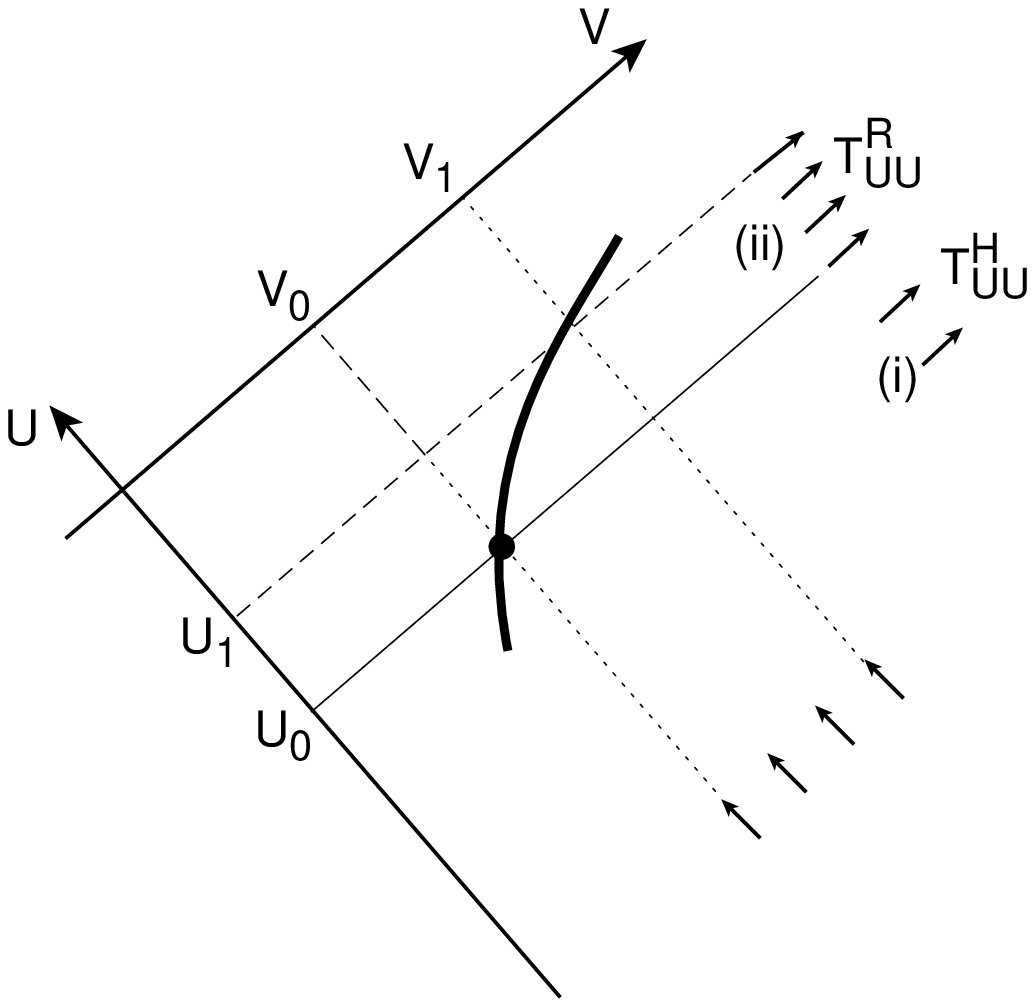}}

Let us first  discuss a situation where the incoming energy density flux is less than $T_{vv}^{\rm cr }$ in the vicinity of the apparent horizon,
  so that it is    time-like (see fig. 1). 
 It may be assumed that this subcritical incident flux is striking on the 
apparent horizon of an already formed black hole.
The  simplest boundary condition is that this low energy-density matter 
   is  just reflected on the time-like apparent horizon. 
A classical reflection on a boundary $V(U)$ is a relation of the form:
$T_{UU}^R=T_{VV} \big({dV\ov dU}\big)^2 $.
 Quantum mechanically, there is
 an additional  contribution, which depends on the normal ordering subtraction
of the composite operators $T_{UU}, T_{VV}$: 
\eqn\ccon{
(T_{UU}^R-t_{UU})= \bigg( {dV\ov dU} \bigg)^2 (T_{VV}- t_{VV} )\ .\ \ \ 
\  }
The calculation given here will not depend on the explicit form of $t_{UU},
t_{VV}$.
In region (i) the outgoing energy-density flux $T_{UU}^H$
can be obtained from the constraint equation,
\eqn\tti{
T_{UU}^{\rm H}= -(4\pi Gr)^{-1}  (\dd_U ^2 r -2 \dd_U \rho  \dd_U r)  +t_{UU}
\ ,\ \ \ \ U<U_0\ .}
For $V\to\infty $ the solution in region (i) approaches the classical Schwarzschild solution,
so that the first term in eq. \tti\   vanishes, and one obtains\foot{
A similar expression  can be derived  from 
the reflection condition \ccon \  which gives
$T^H_{UU}=t_{UU} - \big( {dV\ov dU} \big)^2   t_{VV} $ .
The second term is a small correction to eq. (16) which can be neglected
near a black hole horizon.}
\eqn\tti{
T_{UU}^{\rm H}=  t_{UU}\ ,\ \ \ \ \ U<U_0\ ,
}
which represents the standard Hawking flux.
The total energies radiated in regions (i) and (ii) will be given by 
(we use $4mG dU= -(U+p)du $ ):
\eqn\enei{
E^{\rm (i)}_{\rm out}=4\pi \int _{-\infty }^{u_0} du \ r^2 T_{uu}
=-{\pi\ov mG} \int _{-\infty}^ {U_0} dU (U+p) r^2 T_{UU}^H \ ,
}
\eqn\eneii{
E^{\rm (ii)}_{\rm out}=- {\pi\ov mG} \int _{U_0}^ {U_1} 
dU (U+p) r^2 T_{UU} ^R\ .
}

\ifig\ftwo{ 
Apparent horizon for an   incident flux   greater than   critical.
}{\epsfxsize=5.6cm \epsfysize=5.0cm \epsfbox{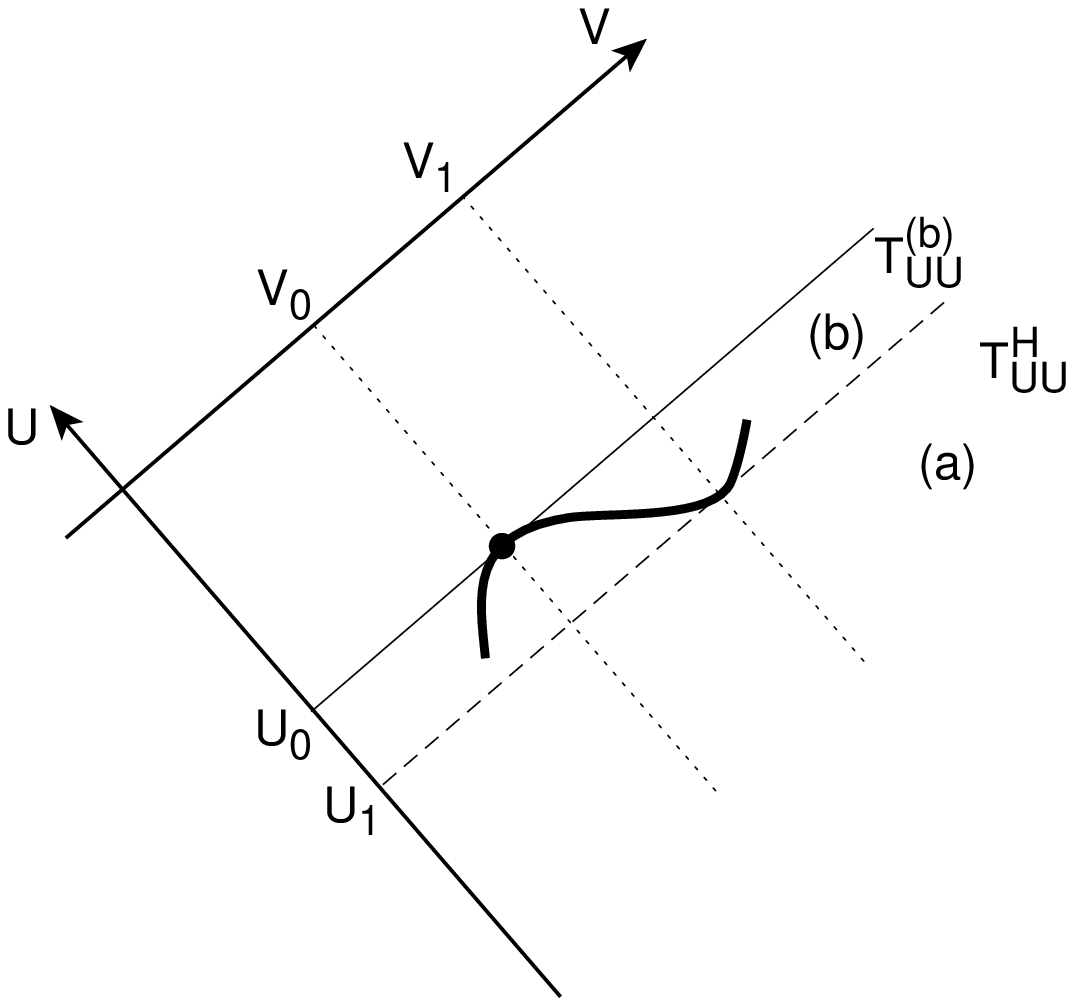}}

\n  Let us now gradually increase $T_{vv}$ above $T_{vv}^{\rm cr }$
so that a part of the apparent horizon becomes space-like, as in fig. 2.
In this process a part of region (i) ends up superposing with region
(ii), giving rise to the  region (b) of fig. 2.
 In this region  the two contributions $T^R_{UU}$
and $T^H_{UU}$ are thus superposed. The correct outgoing $T^{\rm (b)}_{UU}$
can be obtained by carefully continuing the previous formulas.
 Now $U_0 >U_1$, so that   
$$
\int _{-\infty}^{U_0} dU (U+p) T_{UU}^H 
=\int _{-\infty}^{U_1}dU (U+p) T_{UU}^H 
+ \int _{U_1}^{U_0 } dU (U+p) T_{UU}^H \ ,
$$
$$
\int _{U_0}^{U_1}dU (U+p) T_{UU}^R =\int _{U_1}^{U_0}
dU (U+p)(-T^R_{UU})\ .
$$
Therefore, the total energy radiated between $U_1$ and $U_0$  is
\eqn\eneb{
E^{\rm (b)}_{\rm out}=-{\pi \ov mG} \int _{U_1}^{U_0} dU(U+p)
r^2 T^{\rm (b)}_{UU} \ ,\ \ \  
T^{\rm (b)}_{UU}=T^H _{UU}-T^R_{UU}= -
\big( {dV\ov dU} \big)^2 (T_{VV}- t_{VV} ) \ .
} 
Thus  when the apparent horizon is space-like  $T^R_{UU}$ contributes with the reverse sign.
An extra contribution   in region (b) is   not a surprise, since
the geometry in region (b) is   expected to  undergo some modification, 
being in causal contact with the boundary line.
The flip of sign can be physically understood as follows.
  For each given $U'$, the geometry at $V'$
is determined in terms of the energy that has crossed $U'$ at earlier  $V<V'$.
In the presence of the reflecting space-like wall at $U>U_1$, the 
energy-momentum flux crossing $U_1$ cannot be felt by  the geometry in region (b).  
The net effect  is that the geometry in region (b) is changed in such a
way that the flux
$T^R_{UU}$ must be subtracted from the outgoing flux.
This interpretation is confirmed in ref. \amru\  for   an exactly solvable two-dimensional model \rst  , where
the full time-dependent geometry, including the geometry in region (b), can be explicitly obtained.   


It should be noted that only subcritical matter reflects off the (time-like)
apparent horizon. The critical energy density \tcrit\ at the horizon of
a  massive 
black hole is extremely low (e.g. $10^{-64} {\rm g}/{\rm cm}^3 $, for a solar mass black hole).
For infalling objects with energy-density  greater than critical,     
the apparent horizon will be space-like, and they  will just go inside the black hole increasing its mass. As shown below, only a very minor part of their energy and of their information will be emitted. 


\ifig\ftre{
A  macroscopic black hole geometry. The thick line represents the apparent horizon.
}{\epsfxsize=5.5cm \epsfysize=5.2cm \epsfbox{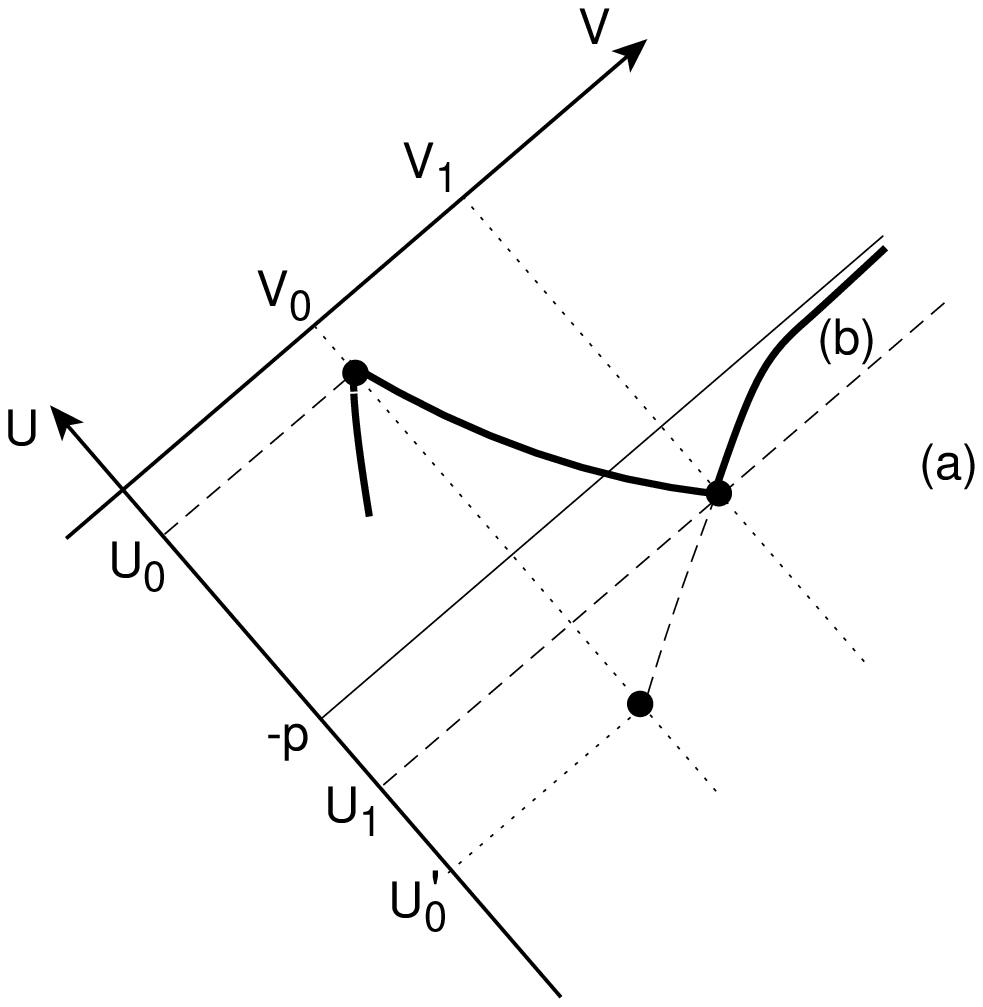}}

Let us   consider the evolution of a  macroscopic black hole geometry,
i.e. with total mass $m\gg m_P, \ m_P=1/\sqrt{ G } $. 
 For convenience we will 
assume that $T_{VV}$ vanishes for $V<V_0$ and $V>V_1$.
 For $V>V_1$ the   geometry is approximately static and given by the Schwarzschild
geometry with   $m=M(V_1)\gg m_P $, $p=P(V_1)\gg |U_0|$.
For $V>V_1$ the apparent horizon curve will be given by the equation
$V(U+p)\cong -kG\ $.
  The geometry is shown in fig.~3. 
Using eq. \eneb , and neglecting
 $t_{VV}$ as compared with the classical collapsing matter contribution $T_{VV}$,  we obtain
\eqn\tttt{
T_{UU}^{\rm (b)}\cong   -T_{VV} \bigg( {dV\ov dU}\bigg) ^2\ .
}
We notice that this outgoing energy-density flux is negative.
It will soon be clear  that the    amount of negative
energy radiated in region (b) is a tiny Planck-scale quantity.
(In quantum theory the energy density 
is not positive definite, and   global energy positivity will not  be violated.
It is the tail of the outgoing wave that carries off
this bit of negative energy).

First, let us estimate the total (positive) energy radiated in region (a).
As  is well known, the Hawking radiation flux is significant only near $U=-p$ 
(more precisely, for $U$ exponentially close to $-p$, $U+p \sim \exp [-\cst Gm^2 ]$~),  where
it has the form 
$T_{uu}^H \sim {1 \ov (Gmr)^2 } $, 
or $T_{UU}^H \sim {1 \ov  r^2 (U+p)^{2} }$,
and it can be neglected for   $U<U'_0\equiv -p-kG/V_0$. Thus 
\eqn\eneas{
E^{\rm (a)}_{\rm out}\cong  -{\pi\ov mG} \int _{U_0'}^{U_1} dU (U+p)
r^2T^H_{UU} \cong 
-{k\ov  4e mG} \log { U_1+p \ov U_0'+p }={k\ov  4e mG} \log { V_1 \ov V_0 }\ ,
}
 which, indeed, is a   small amount of energy. This can be more explicitly seen by
  relating
$\log { V_1 \ov V_0 }$ to the physical parameters characterizing  the incoming energy-density flux, such as the total energy $m$.  
In particular,  consider
   an approximately constant ($v$-independent) flux $T_{vv} $,
which is such that   $T_{vv} \cong \epsi $ at $r\sim 2mG$.
The total mass will be given by 
$m\cong 4\pi r_s^2 \epsi (v_1-v_0)\     \sim 
(mG)^3 \epsi  \log { V_1 \ov V_0 }$. We find
\eqn\eneass{
E^{\rm (a)}_{\rm out}\cong {k\ov 16} (mG)^{-1} {\epsi _{\rm cr }\ov \epsi }\ ,
\ \ \ \ \  \epsi _{\rm cr }\equiv (16\pi e G^3m^2)^{-1}= {\cal T}(V_1) \ .
}
The parameter $ \epsi _{\rm cr } $ is roughly equal to the critical density at which a uniform spherical body would lie within its Schwarzschild radius
(note that $T_{vv}^{\rm cr }$ is much smaller than $ \epsi _{\rm cr }$,
$T_{vv}^{\rm cr }\sim  \epsi _{\rm cr }{m_P^2\ov m^2} $).

Next, we calculate the   (negative) energy received in region (b).
Let $(V_2, U_2\equiv -p)$ be the point at the intersection between the apparent horizon and the
null line $U=-p$, i.e.   $V_2(-p+P(V_2))=-kG $ . 
Let us note that for $m\gg m_P$,   $V_2$ and $V_1$   differ by a small quantity
(it should be remembered that the splitting between the time-like part of the apparent horizon and the    horizon $U=-p$ is a quantum effect). In particular, for a constant density flux one has 
${V_2\ov V_1} \cong  1- {k\epsi _{\rm cr }\ov 16Gm^2 \epsi  } \ $.
The outgoing energy momentum tensor in region (b) is given by eq.~\tttt .
Since we are only interested in the leading order in $m_P/m$, we can 
use  ${dU\ov dV}\cong -P'(V)$.
Inserting   eq. \ppmm\  into eq. \tttt , one obtains 
 \eqn\tttf{
T_{UU}^{\rm (b)} \cong \tt {dV\ov dU}  =
 -  { {\cal T}^2(V) \ov T_{VV}}  \ .
}
$T_{UU}^{\rm (b)}$   carries out   information about the small fraction of the infalling matter that arrived at 
the apparent horizon  between $V_2$ and $V_1$.
In Minkowski coordinates,
\eqn\minkk{
T_{uu}^{\rm (b)}\cong - {V^2 {\cal T}^2(V) \ov (8mG)^2 T_{vv} }
  \exp [-{u\ov 2mG} ] \ .
  }
Hence   $E^{\rm (b)}_{\rm out}\cong 4\pi r^2_s \Delta u  T_{uu}^{\rm (b)}(u_1)
\ ,\ \ \Delta u\sim 2mG $\ ,
\eqn\dido{
E^{\rm (b)}_{\rm out}\cong  -{\pi \epsi  ^2_{\rm cr }\ov 2\epsi } 
mG V_1^2 e^{-{u_1\ov 2mG} } \cong     - a (G^2 m^3)^{-1} \ ,
\ \ \ \ \ a={k^2\epsi  _{\rm cr }\ov 128 e\epsi } <1 \ ,
  }
where we have used $ e^{{u_1\ov 4mG} }=2m V_1/k$.
Thus  the  emitted negative energy  is   smaller than 
$  m_P^4/m^3 $ in absolute value. Since $m\gg m_P$,
   this is a tiny energy (e.g.  for a solar mass black hole,   
$E^{\rm (b)}_{\rm out}\sim - 10^{-114} m_P $).
From eqs. \eneass\ and \dido\ one finds  
 \eqn\didos{
 {|E^{\rm (b)}_{\rm out}| \ov E^{\rm (a)}_{\rm out} }
\cong {m_P^2 \ov m^2} \ll 1 \ .
}
Thus  the total radiated energy
$E^{\rm (a)}_{\rm out} + E^{\rm (b)}_{\rm out}$ is positive
and of order $ E^{\rm (a)}_{\rm out}\sim (mG)^{-1}$ (see eq. \eneass ).

   
To summarize, a simple theory of black hole evolution based on
reflecting boundary conditions 
 on the apparent horizon was described.
 The departure from Hawking theory occurs  precisely
by the time the outgoing modes arise with Planckian frequencies from the vicinity of the horizon (further discussions on the problem of Planck frequencies can be found
in refs. \refs {\svv , \jacobs }). 
The    sudden fall of the subsequent outgoing flux  is caused by a contribution from the expanding trapped surface.
The total radiated energy is a   small  (positive)
Planckian quantity. The final configuration is a stable
black hole geometry, which has retained most of its  mass together with the quantum mechanical information of the original configuration.

 The stability of the final geometry can be understood in different ways.
It is known that in order to have zero fluxes at infinity (in the present case, in region (b)), the gravitational field must be greatly modified near the
line $U=-p$.  This picture is sometimes referred to as the  Boulware vacuum choice, defined  in terms of the  Schwarzschild Killing vector
(here the geometry has settled down to this situation   dynamically   having started from the Unruh vacuum). 
 Accordingly, the  geometry in region (b) will be given by the Schwarzschild metric only at far distances from $U=-p$, viz. for $-V(U+p) \gg \exp [-\cst Gm^2 ]$. 
This condition is satisfied  in the whole of region (b) where $-V(U+p) > kG $, and therefore the corrections to  the  Schwarzschild metric will be exponentially small in the allowed   space-time.
The boundary  does not  imply
that inertial infalling observers will encounter a barrier at the apparent horizon; their description of physics is different (e.g.  
they do not see Hawking radiation) and it may be complementary in the usual
sense of quantum mechanics.


\bigskip

The author wishes to thank D. Amati for useful discussions
and collaboration in the   1+1 dimensional analog \amru ,
and
E. Verlinde for helpful remarks.

\bigskip

\listrefs
\vfill\eject
\end

\bye